# "Faults Unveiled: An In-depth MATLAB Simulation Study of Transmission Line Anomalies"

Milind A. Patwardhan[1], D.P. Chopade[2]

1. Bharati Vidyapeeth's College of Engg. For Women Pune 43(M.S.) INDIA.

2. Bharati Vidyapeeth's College of Engg. For Women Pune 43(M.S.) INDIA.

Correspondence

Milind A.Patwardhan, Bharati Vidyapeeth's College of Engg. For Women Pune 43(M.S.) INDIA.

Email: milind.patwardhan@bharatividyapeeth.edu

**Abstract:**

This study aims to create a MATLAB simulation model to examine three-phase symmetrical and unsymmetrical faults that frequently occur in long transmission line systems. The types of faults considered include single line to ground fault (L-G), double line to ground fault (L-L-G), triple line to ground fault (L-L-L-G), and line to line fault (L-L). The analysis of these faults and their impact on simulation outputs such as voltage and current are investigated. The research utilizes MATLAB software to simulate the transmission line model. The simulation model provides a valuable tool for studying the behaviour of different fault types and understanding their effects on the electrical system. The results obtained from the simulation experiments can aid in developing effective fault detection and protection strategies for transmission lines.

# 1: Introduction:

Fault analysis plays a crucial role in determining the bus voltage and line current when faults occur in a power system. These parameters are of utmost importance in power system analysis [1]. The faults encountered in power systems can be broadly categorized into two types: symmetrical faults and unsymmetrical faults. [2]

A symmetrical fault is a fault where all phases are affected so that the system remains balanced. A three-phase fault is a symmetrical fault. The other three fault types (line to ground, line to line, and two- line to ground) are called unsymmetrical or asymmetrical faults. Analyzing transmission line faults is vital for ensuring the protection and reliability of the power system. To protect the power system, circuit breakers and relay systems are utilized, especially in the case of triple line faults due to their high impact on the transmission system. The simulation of transmission line faults can be efficiently conducted using MATLAB software. The objective of this research work is to study the general types of faults, including balanced and unbalanced faults, in the transmission lines of power systems. Additionally, this research aims to analyze various parameters such as voltage and currents by simulating these types of faults using MATLAB. The simulation results will provide valuable insights into the behaviour and impact of different fault types, enabling the development of effective fault detection and protection strategies for transmission lines.

| S.No | Fault Type | Fault Name | Short Name | Occurrence of fault | Severity of fault |
|---|---|---|---|---|---|
| 1 | Asymmetrical | Line to Ground | L – G | 70 – 80% | Lowest |
| 2 | Asymmetrical | Line to Line | L – L | 15 – 20% | >> |
| 3 | Asymmetrical | Line to Line to Ground | L – L – G | 10% | >> |
| 4 | Symmetrical | Line to Line to Line | L – L – L | 1% | >> |
| 5 | Symmetrical | Line to Line to Line to Ground | L – L – L – G | 2 – 3% | Highest |

*Table 1 Types of faults and their severity*

## 1.1: Symmetrical Fault:

In symmetrical faults, all phases are either short-circuited to each other or to the ground (L-L-L or L-L-L-G). This type of fault is balanced, meaning that the fault currents in each phase are equal in magnitude and are evenly spaced by 120° in phase angle, as illustrated in the figure.

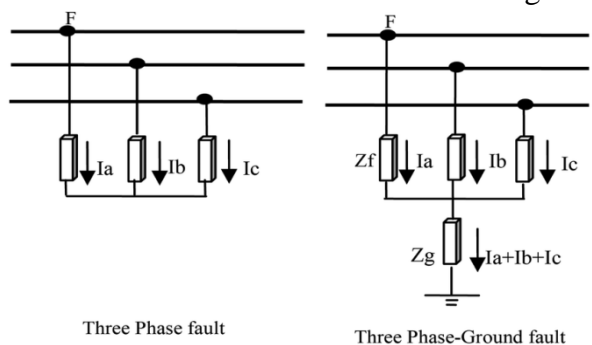

*Figure1: Symmetrical fault*

*Source: https://www.researchgate.net/figure/Two-most-common-symmetrical-fault-types_fig4_3275710/download*

Symmetrical faults, though more severe, are rare. To analyze these large currents, balanced short-circuit calculations are used, requiring only one phase in calculations since the other two phases behave identically due to the fault's balanced nature.

## 1.2: Unsymmetrical Faults:

Unsymmetrical faults affect only one or two phases, causing an imbalance in the three-phase system. These faults lead to unequal line currents with uneven phase angles. Unsymmetrical faults commonly occur between a line and ground or between two lines, and are classified mainly into two types:
- (a) Shunt faults
- (b) Series faults.

Shunt faults include Single Line-to-Ground (L-G), Line-to-Line (L-L), and Line-to-Line-to-Ground (L-L-G) faults, as shown in the figurebelow.

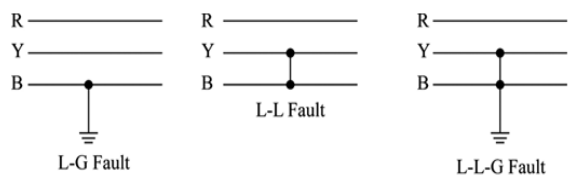

Figure 1: Unsymmetrical fault
Source: Author https://www.chegg.com/homework-help/definitions/unsymmetrical-faults-and-symmetrical-components-4

- Example of Series fault is open conductor circuit.
- These faults mostly occur on power system.
- The calculations of these currents are made by Symmetrical Component method.
- Unbalanced fault analysis is important for Relay setting, single phase switch and system stability studies.

## 2: Methodology:

To model and simulate complex systems, including both linear and nonlinear dynamics, this research utilizes Simulink, a graphical programming extension of MATLAB. Simulink supports multi-domain modeling in continuous, discrete, or mixed-time formats, which provides enhanced flexibility in system design. It features an intuitive Graphical User Interface (GUI), which facilitates the development of sophisticated models and block diagrams through visual programming, thus enabling precise simulation and analysis of system behaviors under varying conditions. [5].

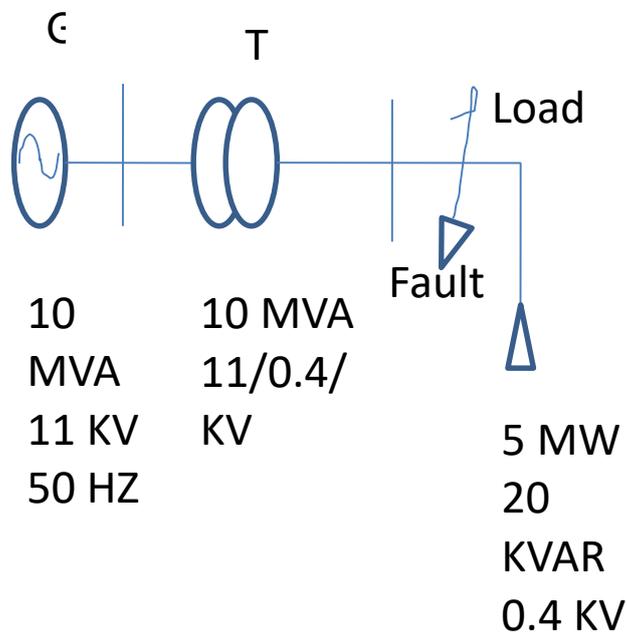

Figure 2 : Power system design parameters
Source: Author

**2.1: Power system design parameters:** The power system is configured with a three-phase generator with a capacity of 10 MVA. The generator operates at a voltage of 11 KV and a frequency of 50 Hz. It is connected to a three-phase transformer with a delta-star configuration, rated at 10 MVA and with a turns ratio of 11/0.4 KV. This transformer is further connected to a load that consumes 5 MW and 20 KVAR at a voltage of 0.4 KV. A three-phase fault is anticipated to occur at the load side.

A standard distribution system is modelled using the predefined parameters presented below.

The voltage and current waveform of simulated faults as shown below are:
a) L-G fault
b) L-L-G fault

c) L-L-L-G fault
d) L-L fault

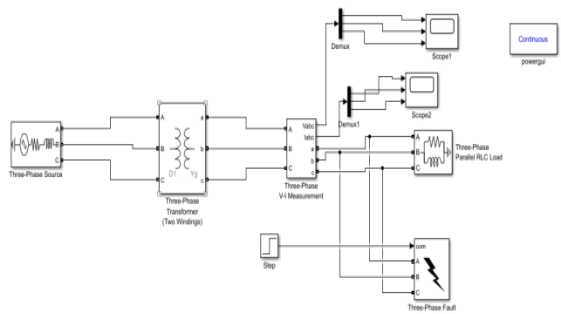

Figure 3: MATLAB simulation model of power system

Different faults are simulated using a model built in MATLAB software, specifically with the Sim Power Systems toolbox. [6]. The representation of this circuit can be observed in Fig.4. The circuit consists of a three-phase source, a three-phase transformer, and a three-phase load. Selecting appropriate ratings for each component, including the three-phase source, transformer, and load, is crucial for accurate simulation results. In order to simulate various fault conditions, a discrete powergui with a sampling time (Ts) of 3e-05 seconds is used in the simulation study. [7].

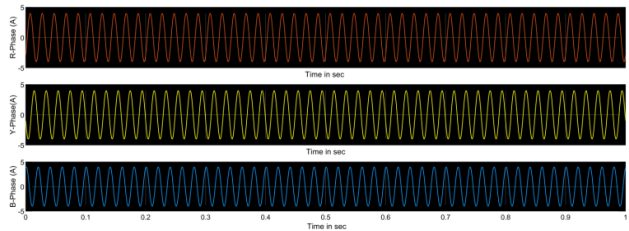

Figure 4 : Healthy system Current waveform

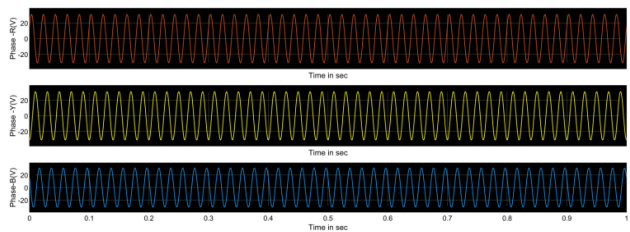

Figure 5: Healthy System Voltage Waveform

## 2.2 Under normal operating conditions:
During normal operation, when no faults are present in any line, all three phases R, Y, B exhibit sinusoidal currents and voltages Fig. 5 and Fig. 6 illustrate the current and voltage waveforms, respectively, of a healthy system. It is important to note that all three phases are balanced in this case.

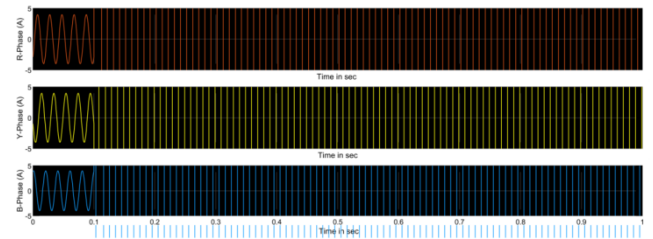

Figure 6: L-L-L-G (with fault) current waveform

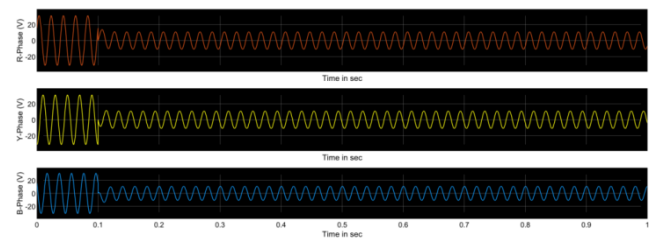

Figure 7: L-L-L-G (with fault) Voltage waveform

## 3: Effect of Fault on Voltage and Current (Load Side)

**3.1.1 Triple line fault:** When a triple line fault occurs in the network, it has a significant impact on the current and voltage waveforms. Fig.7 and Fig.8 illustrate the voltage and current waveforms, respectively, of the system during a triple line fault. These figures demonstrate that at the moment of the fault occurrence at 0.1 sec, the voltages of all three phases R, Y, B abruptly drop, while the current in the system experiences a substantial increase.

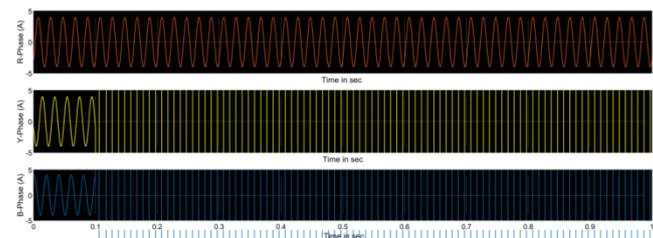

Figure 8: L-L-G (with fault) current waveform

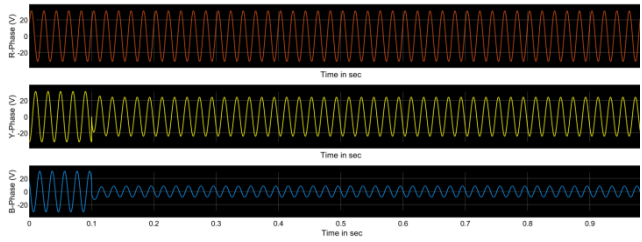

*Figure 9: L-L-G (with fault) voltage waveform*

**3.1.2: Double line to ground fault:** When a double line to ground fault occurs, it has an impact on the current and voltage waveforms. Fig. 9 and Fig. 10 depict the voltage waveforms and current waveforms, respectively, during a double line to ground fault. In these figures, it can be observed that the voltage of two phases Y & B decreases, while the current magnitude of Y & B phases increases.

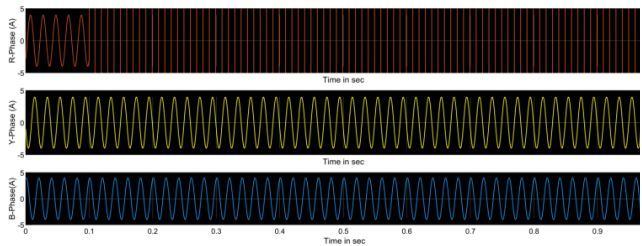

*Figure 10: L-G (with fault) current waveform*

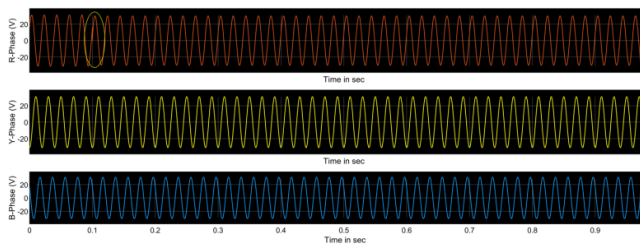

*Figure 11: L-G (with fault) Voltage waveform*

**3.1.3 Single line to ground fault:** During a single line to ground fault, only the voltage of R- phase experiences a drop, while the voltages of the other phases Y and B remain at their initial levels. The current is affected only in the R-phase where the fault occurs, while the currents in the other phases Y&B remain unchanged from their initial values. This behaviour is illustrated in Fig.11 & Fig.12.

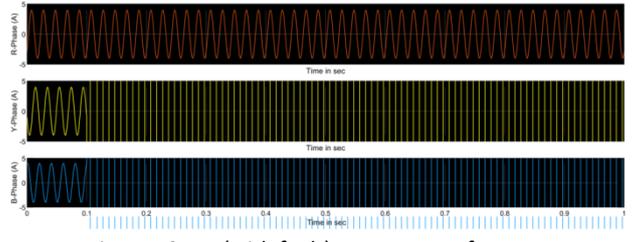

*Figure 12: L-L (with fault) current waveform*

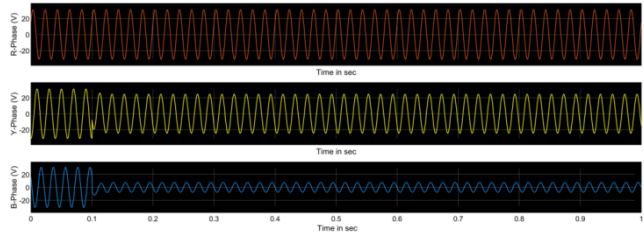

*Figure 13: L-L (with fault) Voltage waveform*

**3.1.4 Line to Line fault**: During a line to line fault, only the voltage of Y and B phase experiences a drop, while the voltage of R phase at its initial level. The current is affected only in the Y and B phase where the fault occurs, while the currents in the R phase remain unchanged from itsinitial value. This behaviour is illustrated in Fig.13 & Fig. 14

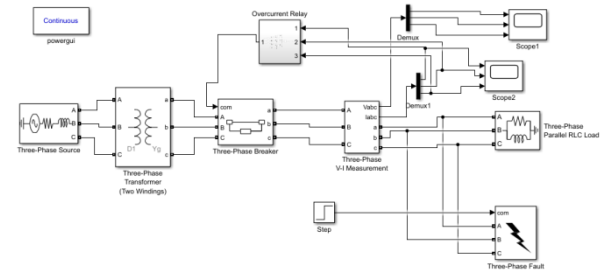

*Figure 14: MATLAB simulation model of power system Three-phase breaker with Relay*

**3.2.1 The MATLAB simulation model of a power system with a three-phase breaker and relay**: The MATLAB simulation model of a power system with a three-phase breaker and relay is designed to analyze and simulate the behaviour of protective relays in response to faults occurring in the system. The purpose of this model is to assess the performance of the breaker and relay system in detecting and isolating faults, ensuring the safety and reliability of the power system.

The simulation model typically consists of the following components:

1. Power System Representation: The power system is represented using appropriate mathematical models in MATLAB. This includes modelling generator, transformer, transmission lines, load, and other relevant components. The parameters of these models are set based on the specific characteristics of the power system being simulated.

2. Three-Phase Breaker: The three-phase breaker [8] is a switching device used to connect or disconnect electrical circuits under normal or fault conditions. In the simulation model, the breaker is represented using appropriate control and switching logic to mimic its behaviour in real-world scenarios. The breaker can be operated manually or automatically, depending on the simulation requirements.

3. Protective Relay: The protective relay is responsible for detecting abnormal conditions, such as faults [9], in the power system and initiating appropriate actions, such as tripping the breaker to isolate the faulty section. The relay in the simulation model is designed to monitor various electrical parameters, such as current, voltage, and frequency, and make decisions based on predefined protection algorithms or logic.

3. Fault Generation: Faults are intentionally introduced into the power system to evaluate the response of the breaker and relay system. Different types of faults, such as L-L-L-G,2-G,L-G& L-L faults can be simulated to assess the effectiveness of the protection scheme. The fault parameters, such as fault location, fault impedance, and fault duration, can be adjusted to simulate different scenarios.

4. Simulation Analysis: The MATLAB simulation model enables the analysis of various aspects related to the performance of the breaker and relay system. This includes evaluating the relay's ability to detect faults accurately and in a timely manner, assessing the breaker's response time, analyzing the coordination between relayin power systems, and studying the impact of different fault conditions on the power system's stability.

By utilizing the MATLAB simulation model of a power system with a three-phase breaker and relay, can gain insights into the protective relay's performance and make improvements to ensure the reliable operation of the power system under both normal and fault conditions.

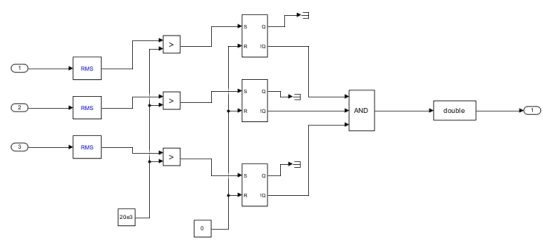

*Figure 15: Relay subsystem*

**3.2.2 The relay subsystem with S-R flip flop**: The relay subsystem with S-R flip flop logic in MATLAB is designed to simulate the behaviour of a protective relay using S-R (Set-Reset) flip flop logic [10]. This relay subsystem is commonly used in power systems to detect fault conditions and initiate appropriate actions, such as tripping a breaker to isolate the faulty section. Here is an overview of how the relay subsystem with S-R flip flop logic can be implemented in MATLAB:

1. Input Signals: The relay subsystem receives input signals from the power system, such as current and voltage measurements. These signals are typically obtained from simulation models or real-time data acquisition systems.

2. Fault Detection Logic: The fault detection logic within the relay subsystem uses the input signals to determine if a fault condition is present. The S-R flip flop logic is applied here to monitor specific conditions and initiate the fault detection process. The S (Set) input is typically triggered by specific fault detection criteria, while the R (Reset) input is activated when the fault condition is cleared [11].

3. Flip Flop State: The S-R flip flop within the relay subsystem maintains its state based on the input signals. When the S input is activated, the flip flop transitions to the "Set" state, indicating the presence of a fault. Conversely, when the R input is activated, the flip flop transitions to the "Reset" state, signifying that the fault condition has been cleared.

4. Trip Signal Generation: Based on the state of the flip flop, the relay subsystem generates a trip signal when the fault condition is detected [12]. This trip signal is used to initiate the necessary protective actions, such as tripping a breaker, to isolate the faulty section of the power system.

5. Simulation and Analysis: The relay subsystem is integrated into the overall power system simulation in MATLAB. The simulation allows for the analysis of the relay's performance, including fault detection time, trip signal generation, coordination with other protective devices, and the overall reliability of the protective scheme. By implementing the relay subsystem with S-R flip flop logic in MATLAB, can be used study the behaviour of the protective relay in a simulated environment. This enables to evaluate and improve the reliability and performance of the protective relay system in power systems.

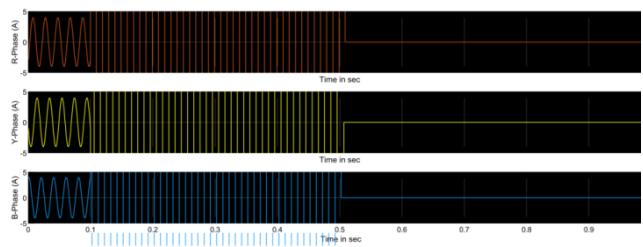

Figure 16: L-L-L-G (fault removed) current waveform

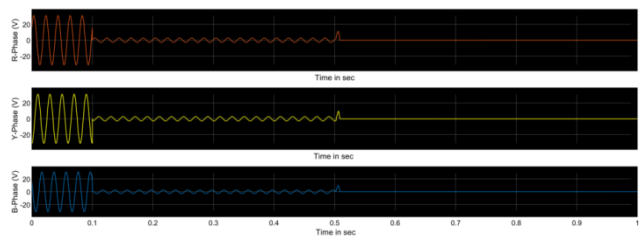

Figure 17: L-L-L-G (fault removed) voltage waveform

**3.3.1 L-L-L-G (fault removed)**: Fig. 17 and fig. 18 shows that, at 0.1 sec, a three-line-to-ground (L-L-L-G) fault occurs in the power system, resulting in a fault current. The fault current waveform exhibits a sudden increase or deviation from the normal operating condition, indicating the presence of the fault. At 0.5 sec, the protective relay detects the fault condition and initiates the activation of the breaker. The breaker is a switching device that is designed to disconnect the faulty section from the rest of the power system. Once the breaker is activated, it opens its contacts, interrupting the flow of current through the faulty section. As a result, the fault current waveform is abruptly interrupted, and the current in the faulty section drops to zero. The voltage waveform may also exhibit a transient response due to the breaker operation, with a brief dip or disturbance in voltage levels during the opening of the breaker contacts. The activation of the breaker isolates the faulty section, preventing further damage and ensuring the safety and stability of the power system. By interrupting the fault current, the breaker protects the system components and allows for the identification and repair of the faulted section.

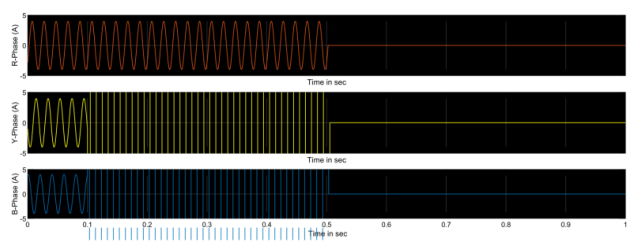

Figure 18: L-L-G (fault removed) current waveform

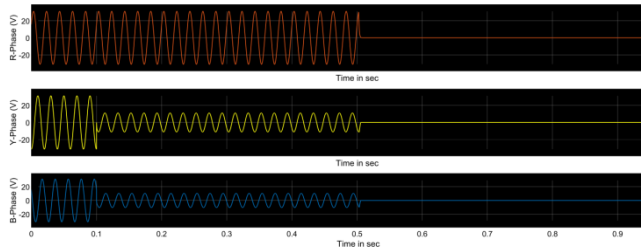

*Figure 19: L-L-G (fault removed) voltage waveform*

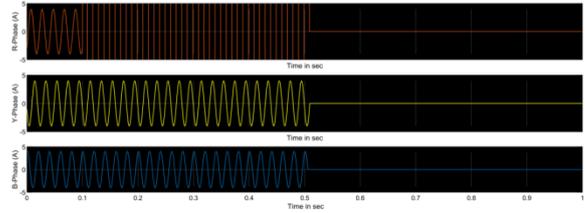

*Figure 20: L-G (fault removed) current waveform*

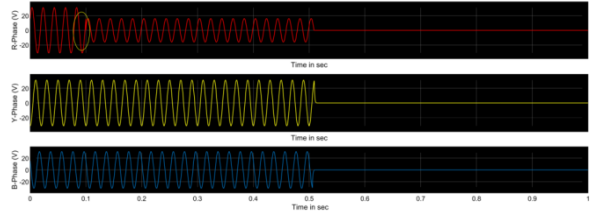

*Figure 21: L-G (fault removed) voltage waveform*

**3.3.2 L-L-G (fault removed):** Fig.19 and Fig. 20 shows that, at 0.1 sec, a two-line-to-ground (L-L-G) fault occurs in the power system, resulting in a fault current. The fault current waveform exhibits a sudden increase or deviation from the normal operating condition, indicating the presence of the fault. At 0.5 sec, the protective relay detects the fault condition and initiates the activation of the breaker. The breaker is a switching device that is designed to disconnect the faulty section from the rest of the power system. Once the breaker is activated, it opens its contacts, interrupting the flow of current through the faulty section. As a result, the fault current waveform is abruptly interrupted, and the current in the faulty section drops to zero. The voltage waveform may also exhibit a transient response due to the breaker operation, with a brief dip or disturbance in voltage levels during the opening of the breaker contacts. The activation of the breaker isolates the faulty section, preventing further damage and ensuring the safety and stability of the power system. By interrupting the fault current, the breaker protects the system components and allows for the identification and repair of the faulted section. It is important to note that a L-L-G fault is a serious fault condition that can lead to significant damage and safety risks in a power system. The timely activation of the breaker is crucial in mitigating the effects of the fault and minimizing the impact on the overall system.

**3.3.3 L-G (fault removed):** Fig. 21 and Fig. 22 show that, at 0.1 sec, a line-to-ground (L-G) fault occurs in the power system, resulting in a fault current. The fault current waveform exhibits a sudden increase or deviation from the normal operating condition, indicating the presence of the fault. At 0.5 sec, the protective relay detects the fault condition and initiates the activation of the breaker. The breaker is a switching device that is designed to disconnect the faulty section from the rest of the power system. Once the breaker is activated, it opens its contacts, interrupting the flow of current through the faulty section. As a result, the fault current waveform is abruptly interrupted, and the current in the faulty section drops to zero. The voltage waveform may also exhibit a transient response due to the breaker operation, with a brief dip or disturbance in voltage levels during the opening of the breaker contacts.The activation of the breaker isolates the faulty section, preventing further damage and ensuring the safety and stability of the power system. By interrupting the fault current, the breaker protects the system components and allows for the identification and repair of the faulted section.The L-G fault is a significant fault condition that can pose serious safety risks and cause damage to equipment. The timely activation of the breaker is crucial in mitigating the effects of the fault and minimizing the impact on the overall system.

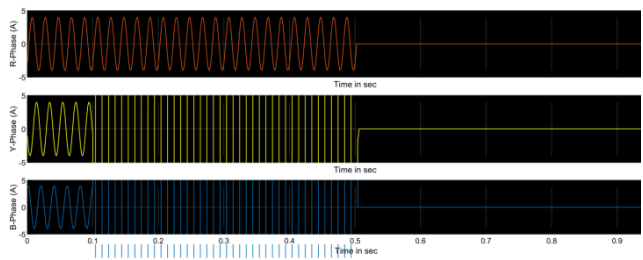
*Figure 22: L-L (fault removed) current waveform*

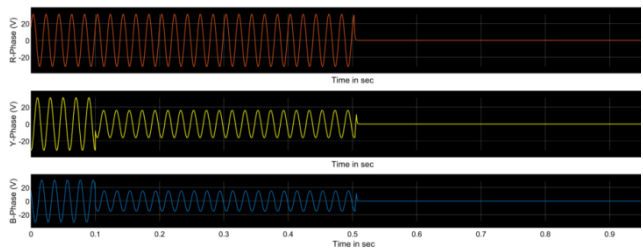
*Figure 23: L-L (fault removed) voltage waveform*

**3.3.4 L-L (fault removed):** Fig.23 and Fig. 24 show that, at 0.1 sec, a line-to-line (L-L) fault occurs in the power system, resulting in a fault current. The fault current waveform exhibits a sudden increase or deviation from the normal operating condition, indicating the presence of the fault. At 0.5 sec, the protective relay detects the fault condition and initiates the activation of the breaker. The breaker is a switching device that is designed to disconnect the faulty section from the rest of the power system. Once the breaker is activated, it opens its contacts, interrupting the flow of current through the faulty section. As a result, the fault current waveform is abruptly interrupted, and the current in the faulty section drops to zero. The voltage waveform may also exhibit a transient response due to the breaker operation, with a brief dip or disturbance in voltage levels during the opening of the breaker contacts. The activation of the breaker isolates the faulty section, preventing further damage and ensuring the safety and stability of the power system. By interrupting the fault current, the breaker protects the system components and allows for the identification and repair of the faulted section. The L-L fault is a significant fault condition that can pose serious safety risks and cause damage to equipment. The timely activation of the breaker is crucial in mitigating the effects of the fault and minimizing the impact on the overall system.

**4. Conclusion:** The MATLAB simulation study provided a comprehensive understanding of the behaviour of fault currents and voltages, allowing for the analysis of their magnitude, duration, and transient responses. The study also examined the performance of protective relays in fault detection, the activation of breakers, and the subsequent isolation of faulty sections to ensure the safety and stability of the power system. By simulating a wide range of fault scenarios, engineers gained insights into fault detection time, trip signal generation, coordination between protective devices, and the overall reliability of the protective scheme. This information is crucial for developing and refining fault detection and protection strategies, leading to enhanced system performance and reduced downtime. The MATLAB simulation study serves as a valuable tool for engineers and researchers in the field of power systems, enabling them to assess the impact of faults on transmission lines and develop effective fault management strategies. It provides a platform to evaluate the performance of protective devices, optimize their settings, and ensure the efficient and reliable operation of the power system. In conclusion, the in-depth MATLAB simulation study of transmission line faults has contributed to our understanding of fault behaviour, protective measures, and system reliability. It serves as a foundation for further research and development in the field of power system protection and helps in designing robust and resilient power systems capable of withstanding and recovering from fault conditions efficiently.